\documentclass[preprint,aps,superscriptaddress,amsmath,amssymb]{revtex4}
\usepackage{bm}
\usepackage{graphicx}

\newcommand{\la}{\lambda}
\newcommand{\ep}{\varepsilon}
\newcommand{\Tr}{{\rm Tr}\,}
\newcommand{\ba}{{\bf a}}

\begin{document}
\title{Quantum control by von Neumann measurements}
\author{Alexander Pechen}\email{apechen@princeton.edu}
\affiliation{Department of Chemistry, Princeton University, Princeton, New Jersey 08544, USA}

\author{Nikolai Il'in}\affiliation{Institute for Problems in Mechanics RAS, prosp.
Vernadskogo 101, Moscow 119526, Russia}

\author{Feng Shuang}\affiliation{Department of Chemistry, Princeton University, Princeton, New Jersey 08544, USA}
\author{Herschel Rabitz}\email{hrabitz@princeton.edu}
\affiliation{Department of Chemistry, Princeton University, Princeton, New Jersey 08544, USA}

\begin{abstract}
A general scheme is presented for controlling quantum
systems using evolution driven by non-selective von
Neumann measurements, with or without an additional
tailored electromagnetic field. As an example, a 2-level
quantum system controlled by non-selective quantum
measurements is considered. The control goal is to find
optimal system observables such that consecutive
non-selective measurement of these observables transforms
the system from a given initial state into a state which
maximizes the expected value of a target operator (the
objective). A complete analytical solution is found
including explicit expressions for the optimal measured
observables and for the maximal objective value given any
target operator, any initial system density matrix, and
any number of measurements. As an illustration, upper
bounds on measurement-induced population transfer between
the ground and the excited states for any number of
measurements are found. The anti-Zeno effect is recovered
in the limit of an infinite number of measurements. In
this limit the system becomes completely controllable. The
results establish the degree of control attainable by a
finite number of measurements.
\end{abstract}
\maketitle

\section{Introduction}
A common goal in quantum control is to maximize the
expected value of a given target operator through
application of an external action to the system. Often
such an external action is realized by a suitable tailored
coherent control field, which steers the system from the
initial state to a target state maximizing the expected
value of the target operator~\cite{R0,R01,R1,R2,R3,R4}. A
coherent field allows for controlled Hamiltonian evolution
of the system. Another form of action on the system could
be realized by tailoring the environment to induce control
through non-unitary system dynamics~\cite{ice}. In this
approach the suitably optimized, generally non-equilibrium
and time dependent distribution function of the
environment (e.g., incoherent radiation or a gas of
electrons, atoms or molecules) is used as a control.
Combining this type of incoherent control by the
environment with a tailored coherent control field allows
for manipulation of both the Hamiltonian and dissipative
aspects of the system dynamics.

Quantum measurements can also be used as an external
action to drive the system evolution towards the desired
control goal. There are two general types of quantum
measurements: instantaneous von Neumann measurements
(selective and non-selective)~\cite{vN} and continuous
measurements~\cite{M}. If the measured operator is
$Q=\sum_i q_i P_i$, where $q_i$ is an eigenvalue of $Q$
with the corresponding projector $P_i$, then the result
$q_i$ of an instantaneous von Neumann measurement of $Q$
is obtained with probability $p_i=\Tr[P_i\rho_{\rm S}]$,
where $\rho_{\rm S}$ is the state of the system just
before the measurement. The state of the system just after
the selective measurement with the result $q_i$ will be
$P_i\rho_{\rm S} P_i/p_i$. If a non-selective measurement
of $Q$ is performed (i.e., if the particular measurement
result is not selected) then the system state just after
the measurement will be $\sum_i P_i\rho_{\rm S}P_i$.

Non-unitary dynamics induced by measurement-driven quantum
evolution was used recently in~\cite{roa1} for mapping an
unknown mixed quantum state onto a known target pure
state. This goal was achieved with the help of sequential
selective measurements of two non-commuting observables.
After each measurement the outcome was observed and used
to decide either to perform the next measurement or to
stop the process. The same problem was studied in the
presence of decoherence introduced by the
environment~\cite{roa2}. The control of the population
branching ratio between two degenerate states by
continuous measurements was considered~\cite{M2}, while
the effect of non-optimized measurements on control by
lasers was investigated~\cite{M1}.

Quantum measurements may be used in a feedback control
scenario~\cite{f0,f1,f2,f3}. In this approach continuous
observations are performed, a controller processes the
results of the measurements, and then based on these
results modifications are made in the coherent control
field in real time to alter the behavior of the quantum
system. Optimal measurements may also be used for quantum
parameter estimation~\cite{qpe1,qpe2}, where the system
state depends on certain $c$-number parameters $\theta_i$.
The goal is to find an optimal measurement strategy to
extract the information about these parameters.

In this paper we explore nonselective von Neumann
measurements to control quantum dynamics. Any measurement
performed on the system during its evolution has an
influence on the dynamics. In particular, non-selective
measurement of an observable with a non-degenerate
spectrum acts on a quantum system by transforming its
density matrix into diagonal form in the basis of
eigenvectors of the operator corresponding to the observed
quantity. Measuring different observables may produce
different changes in the system's state. We optimize the
measured observables such that their consecutive
measurement modifies the density matrix to maximize the
objective. The general formulation includes the use of
optimal measurements along with a tailored coherent
control field. A particular case corresponds to control
only by measurements such that the coherent control field
is not applied. For this case a complete analytical
solution is found for a two-level system. Arbitrary target
operators and initial states of the system are considered.
The solution includes explicit expressions for the optimal
measured observables and the maximal objective values
attained. While control by measurements admits an explicit
analytical solution for two-level systems, the
generalization to the multilevel case is not
straightforward. The situation becomes even more
complicated if coherent control fields are used in
addition to optimized measurements. For this case,
numerical simulations are performed in~\cite{feng} for
several quantum systems controlled by a tailored coherent
control field together with optimization by learning
control of quantum measurements.

The quantum anti-Zeno effect~\cite{BR} can be used to
steer the system from an initial to a target state. In the
anti-Zeno effect continuously measuring the projector
$E(t)=U(t)EU^\dagger(t)$ steers the system into the state
$E(t)$, where $E$ is a projector leaving the initial state
unchanged and $U(t)$ is a unitary operator. Continuous
measurements in the quantum anti-Zeno effect are obtained
as the limit of infinitely frequent von Neumann
measurements. With the anti-Zeno effect the system becomes
completely controllable in this limit. In the laboratory
it may be difficult to perform a large number of
measurements in a short time interval. Thus, a balance may
need to be struck between the number of measurements and
the desired degree of control. In this paper we
analytically establish the degree to which the system can
be controlled by any given finite number of measurements.
This result allows for determining the optimal control
yield in balance with the cost of performing the
measurements.

The paper is organized as follows. In Sec.~II the general
concept of control by measurements is outlined.
Section~III presents the complete analytical solution for
the problem of control by measurements in a two-level
system, and as an application Sec.~IV presents upper
bounds on population transfer by non-selective
measurements. In Sec.~V the relation of this analysis with
the quantum anti-Zeno effect is established. Brief
conclusions are presented in Sec.~VI.

\section{Formulation of control by measurements}
The control "parameters" in the present work are the
observable operators $Q_1,\dots, Q_N$ to be measured. The
number of measurements $N$ can also be optimized if the
cost of each measurement is given. The scheme described
here entails the consecutive laboratory measurement of the
observables $Q_1,\dots,Q_N$ on the same physical system,
but the measurement results are not recorded and not used
for feedback. The latter two restrictions could be lifted,
if desired.

Consider the effect of a non-selective measurement of an
observable $Q$ on the system density matrix. Let
$Q=\sum_iq_iP_i$ be the spectral decomposition of the
observable, where $q_i$ is an eigenvalue and $P_i$ is the
corresponding projector such that $P_i^\dagger=P_i$,
$P_iP_j=\delta_{ij}P_i$ and $\sum_iP_i=\mathbb I$. A
non-selective measurement of the observable $Q$ transforms
the system density matrix $\rho$ into ${\cal
M}_Q(\rho):=\sum_i P_{i}\rho P_{i}$. In particular,
measuring an observable $Q$ which has a non-degenerate
spectrum diagonalizes the density matrix in the basis of
$Q$. In this case ${\cal M}_Q(\rho)=\sum p_{i}P_{i}$,
where $p_{i}=\Tr[\rho P_{i}]$ is the probability to get
the value $q_i$ as the outcome of the measurement.

There are classes of equivalent observables where
measuring an observable $Q$ makes the same transformation
of the system density matrix as measuring any other
observable $\widetilde{Q}$ from the same equivalence
class. Two observables $Q$ and $\widetilde{Q}$ are {\it
measurement-equivalent} if for any density matrix $\rho$
one has ${\cal M}_{Q}(\rho)={\cal
M}_{\widetilde{Q}}(\rho)$. The observables $Q$ and
$\widetilde{Q}$ are measurement-equivalent if their
spectral decompositions have the form $Q=\sum_{i}q_iP_i$
and $\widetilde{Q}=\sum_{i}\tilde q_iP_i$ where for $i\ne
j$ one has $q_i\ne q_j$ and $\tilde q_i\ne \tilde q_j$. In
particular, all observables of the form $Q=q\mathbb I$,
where $\mathbb I$ is the identity operator and $q$ is a
real number, are equivalent to $Q=\mathbb I$. The latter
observables are trivial in the sense that measuring any
such observable does not change the system density matrix.

Let $\rho_0$ be the initial system density matrix.
Consecutively measured observables $Q_1,\dots, Q_N$ modify
the initial system density matrix $\rho_0$ into
\begin{equation}\label{eq3}
\rho_N={\cal M}_{Q_N}\circ{\cal M}_{Q_{N-1}}\dots\circ
{\cal M}_{Q_1}(\rho_0)
\end{equation}
The typical goal in quantum control is to maximize the
expectation value of a target operator $\Theta$ assuming
that initially the system is in a state $\rho_0$. The
objective functional has the form
\begin{equation}\label{eq7}
J_N[Q_1,\dots,Q_N]=\Tr[\rho_N\Theta]
\end{equation}
where $\rho_N$ is defined by~(\ref{eq3}). The control goal
is to find, for given $\rho_0$ and $\Theta$, optimal
observables $Q^{\rm opt}_1,\dots,Q^{\rm opt}_N$ which
maximize the objective functional to produce $J^{\rm
max}_N$.

The general case also includes a tailored coherent
electromagnetic field $\ep(t)$ as a control where the
dynamics of the system is governed by the two forms of
external action: (a) measurements of observables
$Q_1,\dots,Q_N$ at the times $t_1,\dots,t_N$,
respectively, and (b) coherent evolution with a control
field between the measurements. The former action induces
non-unitary dynamics in the system. The latter action
produces unitary evolution of the density matrix $\rho(t)$
between the measurements according to the equation
\begin{equation}\label{eq0}
\frac{d\rho(t)}{dt}=-i[H_0-\mu\ep(t),\rho(t)]
\end{equation}
where $H_0$ is the free Hamiltonian of the system and
$\mu$ its dipole moment. The solution of Eq.~(\ref{eq0})
at a time $t$, with an initial condition $\rho(t_0)=\rho$
at $t_0<t$ is given by a unitary transformation of the
initial density matrix denoted as $U_{[t_0,t)}(\rho)$. In
this notation the system density matrix at a target time
$T\ge t_N$ after $N$ measurements will be
\begin{equation}
\rho(T)=U_{[t_N,T)}\circ{\cal M}_{Q_N}\circ
U_{[t_{N-1},t_N)}\circ{\cal M}_{Q_{N-1}}\circ \dots\circ
U_{[t_1,t_2)}\circ{\cal M}_{Q_1}\circ
U_{[0,t_1)}(\rho_0)\label{eq2}
\end{equation}
The density matrix $\rho(T)$, which is dependent on the
control $\ep(t)$ and $Q_1,\dots Q_N$, determines the
objective functional of the form
$J[\ep(t),Q_1,\dots,Q_N]=\Tr[\rho(T)\Theta]$ with some
target observable $\Theta$. Here, in addition to the
coherent field $\ep(t)$, the observables $Q_1,\dots,Q_N$
are included as variables to be optimized. This general
case is difficult to treat analytically, and for some
models numerical simulations may be performed~\cite{feng}.
In the next section we show that control only by
measurements admits an analytical solution in the case of
a two-level system.

For an atomic multi-level system, practical measurements
of the energy level populations can be performed using
coherent radiation. For example, measuring the population
of energy levels $|1\rangle$ and $|2\rangle$ of a
two-level system can be performed by coupling the ground
or the excited level by a laser pulse to some ancilla
upper level and then observing the spontaneous emission
from the ancilla level to a lower energy level. Such a
measurement is described by projectors
$P_1=|1\rangle\langle 1|$ and $P_2=|2\rangle\langle 2|$
and corresponds to measuring an observable of the form
$Q=q_1P_1+q_2P_2$ with $q_1\ne q_2$. This case shows the
distinction between the use of coherent radiation for
control and for measurements. The coherent radiation used
for controlled unitary evolution generally includes
frequencies close to the transition frequencies of the
controlled energy levels (e.g., levels $|1\rangle$ and
$|2\rangle$ for a two-level system). The radiation used
for measurements includes components with frequencies
close to the transition frequencies between the controlled
levels and the ancilla levels, which moreover should be
subject to decoherence upon decay to the lower energy
levels. In general, measurements also can be performed
through collisions between the system and electrons or
atoms when the scattering cross-section depends on the
initial state of the system. In this case the scattering
data will provide information on the initial state of the
system, thus realizing a measurement procedure.

Measurements on a two-level system in an arbitrary
orthonormal basis $|\psi\rangle$ and $|\psi'\rangle$ can
be realized using an ancilla system and inducing an
interaction Hamiltonian between them, which generates a
unitary evolution operator $U_\psi$ such that for any
vector
$|\chi\rangle=\alpha|\psi\rangle+\beta|\psi'\rangle$ of
the initial system one has
$U_\psi|\chi\rangle|1'\rangle=\alpha|\psi\rangle
|1'\rangle +\beta|\psi'\rangle |2'\rangle$, where
$|1'\rangle$ and $|2'\rangle$ are the energy levels of the
ancilla system. The unitary operator can be chosen as
$U_\psi=\mathbb I-2P$, where $P=|\phi\rangle\langle\phi|$
is the projector onto the one-dimensional subspace of the
composite system spanned by the vector
$|\phi\rangle=|\psi'\rangle(|1'\rangle-|2'\rangle)/\sqrt{2}$.
Then, nonselective measurement of the energy level
populations of the ancilla system [i.e., measurement in
the basis of $|1'\rangle$ and $|2'\rangle$] realizes an
indirect measurement of the initial system in the basis
$|\psi\rangle$ and $|\psi'\rangle$ and changes its state
into $\rho=|\alpha|^2|\psi\rangle\langle
\psi|+|\beta|^2|\psi'\rangle\langle\psi'|$.

An indirect arbitrary von Neumann measurement on some
quantum system can be experimentally realized if the
system can be coupled with another appropriate ancilla
system (the ancilla can be identical to the initial
system) and any unitary operator between these two systems
can be implemented. For the case that the measured system
is a two-level trapped ion, the ancilla could be another
two-level trapped ion. Arbitrary unitary operators between
the two trapped ions can be implemented using a sequence
of at most three controlled-NOT (CNOT) gates and fifteen
elementary one qubit gates~\cite{vatan} (i.e., single ion
unitary operations). Therefore experimental realizations
of CNOT two-qubit gates~\cite{2qubitgates} together with
ability to realize arbitrary one-qubit unitary evolutions
allows for generating any two-qubit unitary operator, in
particular, the operator $U_\psi$ from the preceding
paragraph. Then the ability to measure the ancilla ion in
the energy level basis makes arbitrary measurements on the
initial ion practically possible. The detailed
specification of such a scheme for practical laboratory
realizations of optimal measurements from Sec.~\ref{sec}
requires a separate study.

\section{Control by measurement in a two-level
system}\label{sec} This section presents the analytical
solution for maximizing the expectation value of any given
target observable of a two-level system by optimized
measurements. First, the case with neglect of system free
evolution between the measurements is considered. After
that the modification induced by including the free system
dynamics is described. Any non-trivial observable $Q$ of a
two-level system is an operator in $\mathbb C^2$ with the
form $Q=q_1P_1+q_2P_2$, where $q_1$ and $q_2$ are
eigenvalues and $P_1$ and $P_2$ are the corresponding
projectors such that $P_1P_2=0$ and $P_1+P_2=\mathbb I$.
The observable $Q$ is measurement-equivalent to the
projector $P_1$ (or to $P_2\equiv\mathbb I-P_1$). Thus,
any non-trivial observable of a two-level system is
measurement-equivalent to a suitable projector and the
problem of optimizing over the most general measured
observables is equivalent to optimizing over measurements
of only the projectors.

Any density matrix of a two-level system can be
represented as
\[
\rho=\frac{1}{2}[\mathbb I+{\bf w\cdot\sigma}]
\]
where ${\bf\sigma}=(\sigma_1,\sigma_2,\sigma_3)\equiv
(\sigma_x,\sigma_y,\sigma_z)$ is the vector of Pauli
matrices, ${\bf
w}=(w_1,w_2,w_3)\equiv(w_x,w_y,w_z)\in\mathbb R^3$ is the
Stokes vector, $|{\bf w}|\le 1$, and ${\bf
w}\cdot\sigma=w_x\sigma_x+w_y\sigma_y+w_z\sigma_z$. Thus,
the set of all density matrices of a two-level system can
be identified with the unit ball in $\mathbb R^3$ (i.e.,
the Bloch sphere). Given a density matrix $\rho$, the
components of its Stokes vector can be calculated as
$w_i=\Tr(\rho\sigma_i)$ for $i=1,2,3$.

Pure states correspond to projectors which can be
represented as density matrices with Stokes vectors of
unit norm, $|{\bf w}|=1$. Measuring a projector $P$
transforms the initial density matrix $\rho_0$ into the
new density matrix $\rho_1\equiv {\cal M}_P(\rho_0)$
defined as
\[
\rho_1=P\rho_0 P+(\mathbb I-P)\rho_0(\mathbb
I-P)\equiv\rho_0-[P,[P,\rho_0]]
\]
Let ${\bf a}_0$ and $\bf a$ be the Stokes vectors
characterizing the initial density matrix $\rho_0$ and
projector $P$, respectively (so that $|{\bf a}|=1$). Using
the commutation of Pauli matrices
$[\sigma_k,\sigma_l]=2i\epsilon_{klm}\sigma_k$, where
$\epsilon_{klm}$ is the Levi-Civita symbol, one gets
\begin{eqnarray*}
[P,\rho_0]&=&\frac{1}{4}[(\mathbb I+{\bf
a}\cdot\sigma),(\mathbb I+{\bf a}_0\cdot\sigma)]
=\frac{1}{4}\sum\limits_{k,l=1}^3a_k(a_0)_l[\sigma_k,\sigma_l]\\
&=&\frac{i}{2}\sum\limits_{k,l,m=1}^3\epsilon_{klm}a_k(a_0)_l\sigma_m
=\frac{i}{2}\sum\limits_{m=1}^3({\bf a}\times{\bf
a}_0)_m\sigma_m
\end{eqnarray*}
where $({\bf a}\times{\bf a}_0)$ denotes the vector
product of ${\bf a}$ and ${\bf a}_0$. This gives
\[
\rho_1=\frac{1}{2}[\mathbb I+({\bf a}_0+{\bf a}\times({\bf
a}\times{\bf a}_0))\cdot\sigma]
\]
Using the Lagrange formula $\bf a\times(b\times c) = b(a
\cdot c)-c(a\cdot b)$ for the double vector product and
noticing that $|{\bf a}|=1$ produces
\[
{\bf a}_0+{\bf a}\times({\bf a}\times{\bf a}_0)={\bf
a}_0+{\bf a}({\bf a}\cdot{\bf a}_0)-{\bf a}_0|{\bf
a}|^2={\bf a}({\bf a}\cdot{\bf a}_0)
\]
Therefore we finally have
\[
\rho_1=\frac{1}{2}[\mathbb I+({\bf a}\cdot{\bf a}_0)({\bf
a\cdot\sigma})],
\]
such that the Stokes vector of the density matrix after
measuring $P$ takes the form ${\bf w}_1={\bf a}({\bf
a}\cdot{\bf a}_0)$.

Consider a consecutive measurements of the projectors
$P_1,\dots,P_N$ on the same physical system. Let ${\bf
a}_k$ for $k=1,\dots,N$ be the Stokes vector
characterizing projector $P_k$. After the last measurement
the density matrix will be
\[
\rho_N:={\cal M}_{P_N}\circ {\cal M}_{P_{N-1}}\dots\circ
{\cal M}_{P_1}(\rho_0)=\frac{1}{2}[\mathbb I+{\bf
w}_N\cdot\sigma]
\]
Here the Stokes vector ${\bf w}_N$ has the form
\begin{eqnarray}
{\bf w}_N&=&{\ba}_N(\ba_N\cdot{\bf a}_{N-1})({\bf
a}_{N-1}\cdot{\bf a}_{N-2})\dots({\bf a}_1\cdot{\bf
a}_0)\nonumber\\
&=&{\bf
a_N}\cos\varphi_N\cos\varphi_{N-1}\dots\cos\varphi_1|{\bf
a}_0|\label{eq5}
\end{eqnarray}
where $\varphi_k$ is the angle between vectors ${\bf
a}_{k-1}$ and ${\bf a}_k$.

The objective functional~(\ref{eq7}) may be rewritten as
follows. The target Hermitian operator $\Theta$ can be
represented as $\Theta=\la_0\mathbb
I+{\bm\la}\cdot\sigma$, where $\la_0=\Tr\Theta$ is a real
number and ${\bm\la}=\Tr(\Theta\sigma)\in\mathbb R^3$ is a
three-dimensional real vector. Using this representation
produces $J_N[{\bf a}_1,\dots,{\bf a}_N]=\la_0+{\bf
w}_N\cdot{\bm\la}$. The control parameters are the unit
norm vectors ${\bf a}_1,\dots,{\bf a}_N$ which determine
${\bf w}_N$. Introducing the target vector ${\bf
w}_T={\bm\la}/|{\bm\la}|$ of unit norm, $|{\bf w}_T|=1$,
then the objective functional becomes $J_N[{\bf
a}_1,\dots,{\bf a}_N] =\la_0+|{\bm\la}|({\bf w}_N\cdot{\bf
w}_T)$. It is clear from this expression that maximizing
the objective is equivalent to maximizing the scalar
product ${\bf w}_N\cdot{\bf w}_T$.

Let $\Delta\varphi$ be the angle between ${\bf a}_0$ and
the target vector ${\bf w}_T$ and $\varphi_{N+1}$ be the
angle between vectors ${\bf a}_N$ and ${\bf w}_T$ so that
$\sum_{i=1}^{N+1}\varphi_i\ge\Delta\varphi$. Here the
inequality is used since vectors ${\bf a}_k$ may in
general belong to different planes. The equality may hold
if all vectors ${\bf a}_0,\dots,{\bf a}_N$ and ${\bf w}_T$
belong to the same plane. As a result, the objective
functional may be expressed as
\[
J_N[{\bf a}_1,\dots,{\bf a}_N]=\la_0+|{\bm\la}||{\bf
a}_0|\cos\varphi_{N+1} \cos\varphi_N\dots\cos\varphi_1
\]
The objective $J_N$ is maximized if
$\varphi_1=\varphi_2=\dots=\varphi_N=
\varphi_{N+1}=\Delta\varphi/(N+1)$ so that
$\sum_{i=1}^{N+1}\varphi_i=\Delta\varphi$, and the maximal
value of the objective is
\begin{equation}\label{eq4}
J_N^{\rm max}=\la_0+|{\bm\la}||{\bf a}_0|
\Bigl[\cos\frac{\Delta\varphi}{N+1}\Bigr]^{N+1}
\end{equation}
The corresponding optimal $k$-th measured observables are
those which are measurement-equivalent to the projector
$P^{\rm opt}_k=\frac{1}{2}[\mathbb I+{\bf a}^{\rm
opt}_k\cdot\sigma]$, where the vector ${\bf a}^{\rm
opt}_k$ belongs to the plane formed by ${\bf a}_0$ and
${\bf w}_T$ and is obtained by rotating the unit norm
vector ${\bf a}_0/|{\bf a}_0|$ by the angle
$k\Delta\varphi/(N+1)$. Any such observable has the form
\begin{equation}\label{eq8}
Q^{\rm opt}_k=q_kP^{\rm opt}_k+\tilde q_k(\mathbb I-P^{\rm
opt}_k)
\end{equation}
where $q_k$ and $\tilde q_k$ are real numbers with $q_k\ne
\tilde q_k$. It is not important for the control purposes
here which observables from the equivalence class are
chosen. In particular, the projector $P^{\rm opt}_k$ could
be used as the measured observable. In general, if the
free system evolution is neglected then all the vectors
characterizing optimal observables must belong to the same
plane formed by the Stokes vectors of the initial and
final states.

The Stokes vector of the system density matrix after
measuring $P^{\rm opt}_k$, in the case that the total
number of measurements is $N$, will be
\begin{equation}\label{eq6} {\bf
w}_{N,k}= |{\bf a}_0|
\Bigl[\cos\frac{\Delta\varphi}{N+1}\Bigr]^k {\bf a}^{\rm
opt}_k.
\end{equation}
Thus, each optimal measurement rotates the Stokes vector
of the density matrix by the angle $\Delta\varphi/(N+1)$
and shortens its length by the factor
$\cos[\Delta\varphi/(N+1)]$.

The analysis above assumes that the free system evolution
between the measurements could be neglected. This limit
will be valid if the time between measurements $\Delta t$
is sufficiently small such that $\Delta t E_i\ll1$ for the
eigenvalues $E_i$, $i=1,2,\dots$ of $H_0$; the limit is
also valid if the system is close to degenerate,
$E_i\approx E_j$, $\forall i,j$. In order to go beyond
this limiting case, now we will describe the modification
induced by the free dynamics. Suppose that $N$
measurements of observables
$\widetilde{Q}_1,\dots,\widetilde{Q}_N$ are performed at
the fixed time moments $0\le t_1<t_2<\dots<t_N\le T$ and
between the measurements the system evolves with its time
independent free Hamiltonian $H_0$. Then the density
matrix $\rho(T)$ at the target time $T$ is given by the
equation of the form~(\ref{eq2}) with the unitary
evolution between $(k-1)$-th and $k$-th measurements
$U_{[t_{k-1},t_k)}(\rho) =\exp[-i(t_k-t_{k-1})H_0]\rho
\exp[i(t_k-t_{k-1})H_0]$. The relation ${\cal
M}_{\widetilde{Q}}(U\rho U^\dagger)=U{\cal
M}_{Q}(\rho)U^\dagger$ with $Q=U^\dagger \widetilde{Q}U$
gives by induction $\rho(T)=e^{-iTH_0}\rho_N e^{iTH_0}$,
where $\rho_N={\cal M}_{Q_N}\circ\cdots\circ{\cal
M}_{Q_1}(\rho_0)$ is the density matrix evolved only under
the measurements of the modified operators
$Q_k=e^{it_kH_0} \widetilde{Q}_ke^{-it_kH_0}$. Therefore
the objective function for a target operator
$\widetilde{\Theta}$ in the case of including the free
dynamics,
$\widetilde{J}_N[\widetilde{Q}_1,\dots,\widetilde{Q}_N;\widetilde{\Theta}]
:=\Tr[\rho(T)\widetilde{\Theta}]$, equals the objective
function $J_N[Q_1,\dots,Q_N;\Theta]:=\Tr[\rho_N\Theta]$
without the free dynamics with the modified target
operator $\Theta=e^{iTH_0}\widetilde{\Theta} e^{-iTH_0}$
and measured observables
$Q_k=e^{it_kH_0}\widetilde{Q}_ke^{-it_kH_0}$. The latter
problem was completely solved above and the optimal
measured observables are given by~(\ref{eq8}). If $Q^{\rm
opt}_k$ are such optimal measured operators for the
objective function $J_N[Q_1,\dots,Q_N;\Theta]$ with
neglected free dynamics, then the optimal measured
operators for the objective function
$\widetilde{J}_N[\widetilde{Q}_1,\dots,\widetilde{Q}_N;\widetilde{\Theta}]$
are $\widetilde{Q}^{\rm opt}_k=e^{-it_kH_0}Q^{\rm
opt}_ke^{it_kH_0}$. This implies that, while all the
vectors $\ba^{\rm opt}_k$ corresponding to the optimal
operators $Q^{\rm opt}_k$ belong to the same plane, this
is not true for the vectors $\widetilde{\ba}^{\rm opt}_k$
corresponding to the operators $\widetilde{Q}^{\rm
opt}_k$.

Thus, if the free evolution between the measurements is
not important then, for arbitrary initial and target
states, the vectors characterizing optimal measured
observables must belong to the same plane formed by the
Stokes vectors of the initial and final density matrices.
If the free evolution is relevant, then the optimal
observables undergo additional unitary transformations
with the generator $H_0$, which moves their corresponding
vectors out of a plane.

\section{Measurement-induced population transfer}\label{sec2}
\begin{figure*}[p]\center
\includegraphics[scale=0.93]{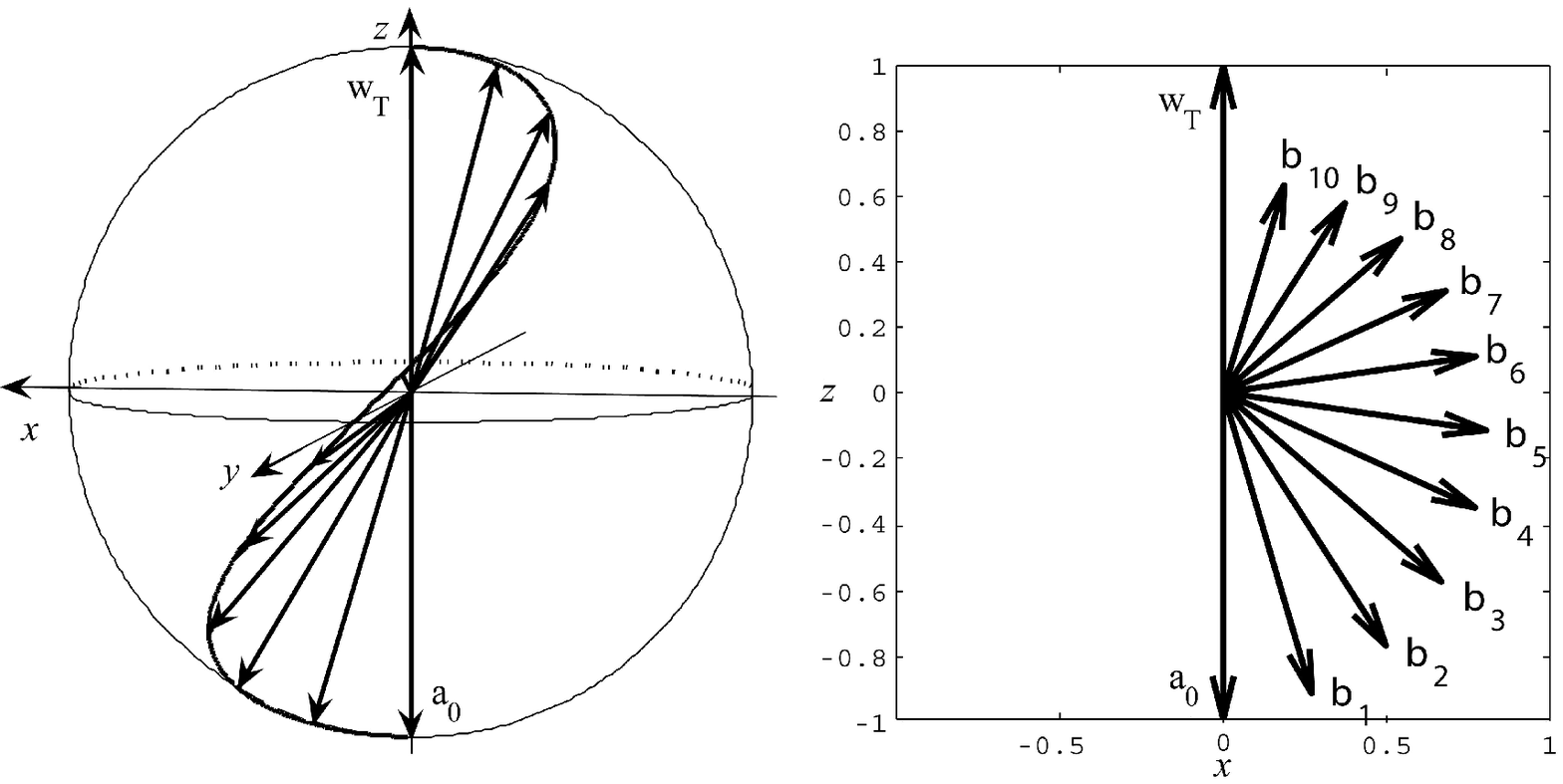}
\caption{The plots illustrate evolution of a two-level
system during the optimal measurement-induced population
transfer from the ground to the excited state. The vectors
${\bf a}_0=-e_z$ and ${\bf w}_{\rm T}=e_z$ are the Stokes
vectors of the initial and target states. The total number
of measurements is $N=10$. The left plot shows the ten
unit norm vectors describing the optimal measured
observables for a case with non-trivial free system
dynamics. The smooth curve passing through the ends of
these vectors describes the projectors for the anti-Zeno
effect in the limit of an infinite number of optimal
measurements. The right plot shows the Stokes vector ${\bf
b}_{k}\equiv{\bf w}_{10,k}$, defined by~(\ref{eq6}), of
the system state after $k$ optimal measurements with
neglected free dynamics. Each measurement rotates the
preceding vector by the angle $\varphi_1=\pi/11$
anticlockwise and shortens its length by the factor
$\cos\varphi_1$.}\label{fig22}
\end{figure*}

\begin{figure*}[p]\center
\includegraphics[scale=1]{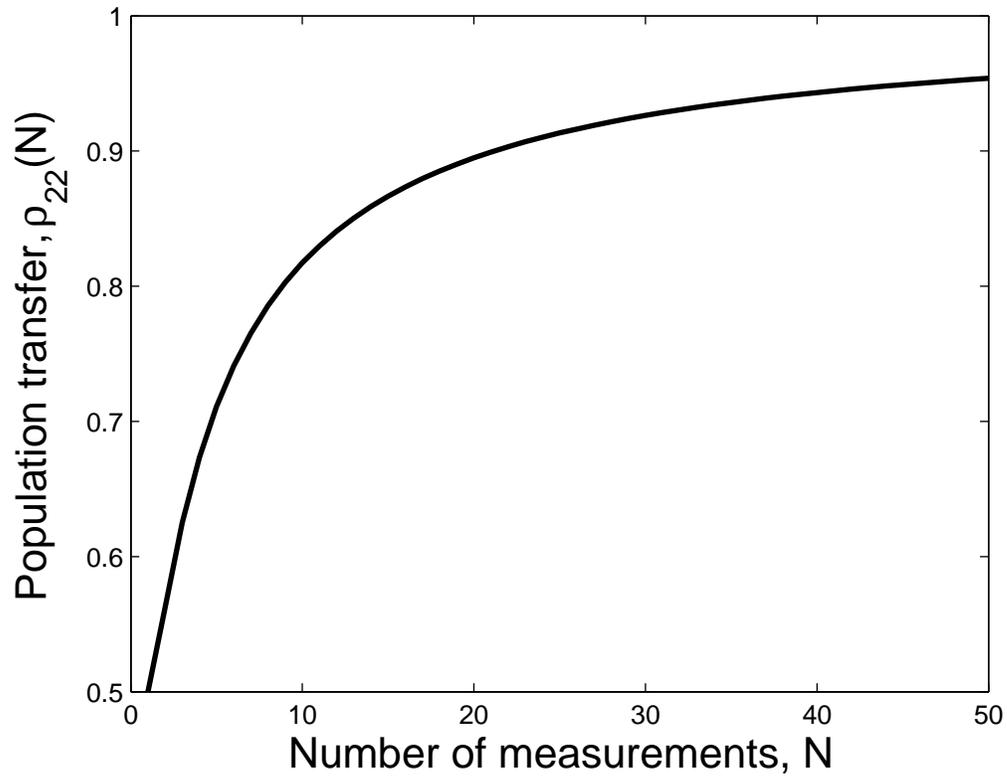}
\caption{Maximal measurement-induced population transfer
from the ground to the excited state of a two-level system
as a function of the number of optimal measurements. The
maximal population $\rho_{22}(N)$ approaches $1$ as
$N\to\infty$.}\label{fig21}
\end{figure*}

Here we apply the general result~(\ref{eq4}) to the
problem of population transfer between orthogonal states
$|1\rangle$ and $|2\rangle$ of a two-level system. The
initial density matrix is $\rho_0=|1\rangle\langle 1|$.
The target operator is the projector on the excited state,
$\Theta=|2\rangle\langle 2|$, which corresponds to
$\la_0=1/2$ and ${\bm\la}=1/2e_z$. In this case ${\bf
a}_0=-e_z$, ${\bf w}_T=e_z$ and therefore the angle
between the initial and target vectors is
$\Delta\varphi=\pi$ (see Fig.~\ref{fig22}). The maximal
population transfer to the excited level $\rho_{22}(N)$ as
a function of the number of performed measurements is
given by~(\ref{eq4}) and has the form
\begin{equation}\label{eq10}
\rho_{22}(N)=\frac{1}{2}
\Bigl[1+\Bigl(\cos\frac{\pi}{N+1}\Bigr)^{N+1}\Bigr]
\end{equation}
The function $\rho_{22}(N)$ for $N\le 50$ is plotted in
Fig.~\ref{fig21}.

Suppose that the goal is to transfer population
$\rho_{22}=1-\ep$ to the excited level, where $\ep\ll 1$.
The asymptotic number of optimal measurements necessary to
meet this goal may be found as follows:
\begin{eqnarray*}
\rho_{22}(N)&=&1-\ep\Rightarrow
1+\Bigl[\cos\frac{\pi}{N+1}\Bigr]^{N+1}=2-2\ep\\
&\Rightarrow&
\cos\frac{\pi}{N+1}=(1-2\ep)^{1/(N+1)}\approx1-\frac{2\ep}{N+1}
\end{eqnarray*}
Notice that a small value of $\ep$ requires $N$ to be
large. Therefore the Taylor expansion for
$\cos(\pi/(N+1))$ may be used which gives the approximate
relation
\[
\frac{\pi^2}{2(N+1)^2}\approx\frac{2\ep}{N+1}
\]
It then follows that the number of measurements necessary
to transfer population $\rho_{22}=1-\ep$ to the excited
level asymptotically behaves as $N\approx\pi^2/(4\ep)$,
which is consistent with the behavior in Fig.~\ref{fig21}
with $N=50$ and $\ep\simeq 0.05$.

Figure~\ref{fig22} illustrates the population transfer by
optimal measurements in a two-level system for $N=10$. The
left-hand plot corresponds to the case with a non-trivial
free dynamics driven by the free Hamiltonian
$H_0=\frac{1}{2}\sigma_z$. The 10 observables
$\widetilde{Q}_1,\dots,\widetilde{Q}_{10}$ are measured at
the time moments $t_k=Tk/(N+1)$, where the target time is
chosen as $T=\pi$, and characterized by the unit norm
vectors
\begin{equation}
\widetilde{{\bf a}}^{\rm opt}_k=-\cos\Bigl(\frac{\pi
k}{N+1}\Bigr)e_z+\sin\Bigl(\frac{\pi k}{N+1}\Bigr)
\Bigl[\cos\Bigl(\frac{Tk}{N+1}\Bigr)e_x
+\sin\Bigl(\frac{Tk}{N+1}\Bigr)e_y\Bigr],\quad
k=1,\dots,N\label{eq9}
\end{equation}
shown on the left-hand plot. The smooth curve passing
through the ends of these vectors represents the
continuous family of the projectors characterizing the
anti-Zeno effect in the limit of infinite number of
optimal measurements, as described in the next section.

The right-hand plot in Fig.~\ref{fig22} illustrates the
evolution of the system without free dynamics between the
measurements. In this case
$\varphi_1=\dots=\varphi_{11}=\pi/11$ and the optimal
observable for the $k$-th measurement is characterized by
the vector
\begin{equation}\label{eq9v2}
{\bf a}^{\rm opt}_k= -\cos\Bigl(\frac{\pi k}{N+1}\Bigr)
e_z+\sin\Bigl(\frac{\pi k}{N+1}\Bigr)e_x,
\end{equation}
which is obtained by rotating the initial vector ${\bf
a}_0$ by the angle $k\varphi_1$. The system density matrix
after the $k$-th measurement has the Stokes vector ${\bf
b}_k\equiv{\bf w}_{10,k}=(\cos\varphi_1)^k{\bf a}^{\rm
opt}_k$. The plot shows the length of ${\bf b}_k$
decreasing after each measurement by the factor
$\cos\varphi_1$. The relation between the two cases is
that each vector $\widetilde{{\bf a}}^{\rm opt}_k$
describing optimal $k$-th measurement for the case with
free dynamics is obtained by rotating ${\bf a}^{\rm
opt}_k$ by the angle $\pi k/10$ around $z$-axis. Note that
the ten vectors on the left plot of Fig.~\ref{fig22} are
the vectors $\widetilde{{\bf a}}^{\rm opt}_k$
characterizing the ten optimal measured observables which
are measurement-equivalent to the projectors, i.e., pure
states, $\widetilde{P}^{\rm opt}_k=\frac{1}{2}[\mathbb
I+\widetilde{{\bf a}}^{\rm opt}_k\cdot\sigma]$ and
therefore these vectors have unit norm. The ten vectors
${\bf b}_k$ on the right plot characterize the system
density matrix $\rho=\frac{1}{2}[\mathbb I+{\bf
b}_k\cdot\sigma]$ after each measurement, which is a mixed
state due to decoherence induced by the measurements, and
therefore these vectors have norm less than one. The
shortening of the vectors characterizing the system
density matrix after each measurement for the case
illustrated on the left plot will be the same, as for the
case shown on the right plot, i.e., this shortening is by
the factor $\cos\varphi_1$ after each measurement.

Numerical simulations for some models in~\cite{feng} also
suggest that the expression~(\ref{eq10}) gives the
estimate for the maximal population transfer between any
pair of orthogonal levels in any multi-level quantum
system. The complete analytical investigation of this
problem remains open for a future research.

\section{Relation with the anti-Zeno effect}
The maximal population transfer to the excited level by a
finite number of measurements satisfies $\rho_{22}(N)<1$.
Since $\lim\limits_{n\to\infty}[\cos(\pi/n)]^n=1$, one has
$\lim\limits_{N\to\infty}\rho_{22}(N)=1$, i.e., complete
population transfer is attained only in the limit of an
infinite number of measurements which describes the
anti-Zeno effect. To show this behavior consider the
projector $E=|1\rangle\langle 1|=\frac{1}{2}[\mathbb
I+{\bf a}_0\cdot\sigma]$. The goal is to steer the system
from the initial state $|1\rangle$ at time $t=0$ into the
target state $|2\rangle$ at time $t=T$. Define for each
$t\in[0,T]$ the unitary operator $U(t)=\exp(i\sigma_y\pi
t/2T)$. Then $E(t):=U(t)EU^\dagger(t)$ is the projector
characterized by the Stokes vector ${\bf w}_t=-\cos(\pi
t/T)e_z+\sin(\pi t/T)e_x$, i.e., $E(t)=\frac{1}{2}[\mathbb
I+{\bf w}_t\cdot\sigma]$. According to the anti-Zeno
effect, continuous measurement of the projector $E(t)$
steers the system at time $t$ into the state characterized
by vector ${\bf w}_t$. One has ${\bf w}_{\rm T}=e_z$,
i.e., at the target time the system will be transferred
into the state $|2\rangle$.

The anti-Zeno effect is obtained in the limit of an
infinite number of measurements when the interval between
any two consecutive measurements tends to zero. Consider
performing $N$ measurements of the optimal observables
$Q^{\rm opt}_1,\dots,Q^{\rm opt}_N$, where $Q^{\rm
opt}_k=\frac{1}{2}[\mathbb I+\ba ^{\rm opt}_k\cdot\sigma]$
and $\ba^{\rm opt}_k$ is defined by Eq.~(\ref{eq9v2}). If
the dynamics between the measurements is neglected then
the state of the system after the $k$-th measurement will
be characterized by the vector
\[
{\bf w}_{N,k}= \Bigl[\cos\frac{\pi}{N+1}\Bigr]^k
\Bigl[-\cos\Bigl(\frac{\pi k}{N+1}\Bigr)e_z
+\sin\Bigl(\frac{\pi k}{N+1}\Bigr)e_x\Bigr]
\]
Taking the limit as $N,k\to\infty$ such that the ratio
$k/(N+1)=t/T$ is kept fixed produces
\[
\lim\limits_{N,k\to\infty}{\bf w}_{N,k}=-\cos(\pi
t/T)e_z+\sin(\pi t/T)e_x\equiv {\bf w}_t
\]
Thus, the anti-Zeno effect is recovered in this limit. The
corresponding evolution is characterized by the projector
$E(t)$ describing the rotation of the Stokes vector of the
system density matrix in the same plane. In general, other
evolutions exist which steer the system from the ground to
the excited state with projectors $\widetilde{E}(t)$ which
induce rotations out of a plane. Such projectors can be
limits of optimal control by a finite number of
measurements if the free evolution between the
measurements is non-trivial. As an example of this
situation, the smooth curve on the left plot of
Fig.~\ref{fig22} describes the optimal anti-Zeno effect
characterized by the projector
$\widetilde{E}(t)=\frac{1}{2}[\mathbb I+\widetilde{\bf
w}_t\cdot\sigma]$, where
\begin{equation*}
\widetilde{\bf w}_t=
\lim\limits_{N,k\to\infty}\widetilde{\ba}^{\rm opt}_k
=-\cos(\pi t/T)e_z+\sin(\pi t/T)[\cos(\pi t/T)e_x+\sin(\pi
t/T)e_y]
\end{equation*}
Here the vectors $\widetilde{\ba}^{\rm opt}_k$ defined
by~(\ref{eq9}) characterize the optimal measurements for
the example with the free evolution considered in
Sec.~\ref{sec2}, the target time is chosen as $T=\pi$, and
the limit is taken with fixed ratio $k/(N+1)=t/T$.

\section{Conclusions}
In this paper control of quantum systems by non-selective
measurements is considered. The capabilities of optimized
measurements for control of a two-level system are
explicitly investigated. The optimal observables and
maximal expectation value of any target observable are
analytically found given any initial system density matrix
and any fixed number of performed measurements, thus
providing a complete analytical solution for control by
measurements in a two-level system. For any given number
of measurements the degree of control, i.e., the maximum
value of the objective, is found. The relation between the
optimal measurements and the quantum anti-Zeno effect is
established. Looking ahead, the ultimate goal will be
specification of laboratory protocols to make the
procedure of control by measurement practical for
realistic systems.

\begin{acknowledgments}
Three of the authors (A.P., F.S., and H.R.) acknowledge
support from the National Science Foundation.
\end{acknowledgments}

\end{document}